# Structural and magnetic ordering in $Pr_{0.65}(Ca_ySr_{1-y})_{0.35}MnO_3$: "quantum critical point" versus phase segregation scenarios.[†]

G. R. Blake[a,b*], L. Chapon[c*], P. G. Radaelli[a,d], D. N. Argyriou[c], M. J. Gutmann[a] and J. F. Mitchell[b]

[a] ISIS Facility, Rutherford Appleton Laboratory-CCLRC, Chilton, Didcot, Oxfordshire, OX11 0QX, United Kingdom

[b] Materials Science Division, Argonne National Laboratory, Argonne, IL 60439, USA

[c] Hahn-Meitner-Institut, Glienicker Straße 100, 14109 Berlin (Wannsee), Germany

[d] Department of Physics and Astronomy, University College London, Gower Street, London, WC1E 6BT, United Kingdom.

## ABSTRACT

The phase diagram of $Pr_{0.65}(Ca_ySr_{1-y})_{0.35}MnO_3$, $0.6 \leq y \leq 0.8$, has been determined by neutron diffraction, magnetization and electrical conductivity measurements in order to investigate the nature of the transition between ferromagnetic metallic and charge-ordered insulating states near y=0.75. Two possible scenarios for this transition have been proposed: a "quantum critical point"-like feature, near which an associated

---

charge-disordered paramagnetic phase is present, or a phase coexistence region. We demonstrate that the latter case is true, phase segregation occurring on a mesoscopic/macroscopic length-scale (several hundred Ångstroms to several microns). Our results show that no significant amount of the charge-disordered paramagnetic phase is present at low temperatures. Our data also indicate that the charge-ordered insulator to ferromagnetic metal phase boundary is temperature as well as composition-dependent.

## 1. Introduction

There is consensus on the statement that different parts of the manganite perovskite phase diagram have distinct phenomenology, which can only be understood by using different "ingredients" to construct their physical picture. This assertion, whilst definitely true across the whole multi-dimensional phase diagram, is put in strong focus if one considers how properties change as a function of the tolerance factor[1]. By changing the size of the $A$-site ionic radius at constant doping in $A$MnO$_3$, one is able to affect the Mn-O-Mn bond angles, which is expected to result in rather modest changes of the one-electron bandwidth. However, by tuning this parameter near optimum doping, one is able to produce materials with completely different properties, varying from charge- and orbital-ordered (CO) insulator to ferromagnetic (FM) metal.

The properties of the higher $T_C$, rhombohedral compounds such as La$_{2/3}$Sr$_{1/3}$MnO$_3$ can be well understood in the framework of the conventional double-exchange (DE) mechanism, although detailed agreement with the experiments can only be reached using modern theoretical and computational tools[2]. As the electronic bandwidth is reduced at constant doping, the crystallographic symmetry is lowered to orthorhombic, thereby favoring the formation of Jahn-Teller (JT) polarons. As a consequence, the Curie temperature $T_C$ is suppressed well below the value predicted by DE theory. The intrinsic inhomogeneities occurring in this part of the phase diagram have been evidenced by a variety of techniques, sensitive to short-range (0.5-2 nm) FM ordering[3] or JT polaron correlations[4]. Also, several studies have demonstrated that percolation between insulating and conducting domains plays a critical role in determining the electrical conductivity of these materials[5]. Nevertheless, there have been many attempts to treat the problem in the framework of an "effective medium" approximation. Early on, Millis[6, 7] proposed a model in which electron-phonon coupling favored the formation of isolated polarons in the paramagnetic state. As details of the nanostructure of these compounds became better known, more complex theoretical and computational models were proposed[8]. Recently, the emphasis has been on including the role of chemical disorder in nucleating relatively stable JT-correlated clusters, which would be gradually suppressed on cooling. In this spirit, a good quantitative account of the magnetic, transport and thermodynamic properties has been recently given by Salamon *et al.*[9] in terms of the so-called "Griffiths phase" treatment.

As the bandwidth is further reduced, one encounters the boundary between the FM, metallic phase and an antiferromagnetic (AFM) insulator with partial charge and orbital ordering [1, 10]. Here, the magnetic and transport properties can be strongly affected by external parameters, such as magnetic field (magnetoresistive ratios up to 11 orders of magnitude have been reported[11]), pressure[12] and illumination with different kinds of radiation [13]. In addition, phase coexistence over much longer length-scales (from sub-micron to several microns) has been reported[14-16]. Caution should therefore be exercised in extending "effective medium" approximations to this regime. Nevertheless, in a recent publication, Burgy *et al.*[17] present extensive computational investigations on a "toy model", specifically constructed to capture the phenomenology of this part of the phase diagram. This Ising-like system contains the basic ingredients believed to be relevant to the physics, that is, competing interactions (giving rise, as a function of a coupling constant $J_2$ between next-near neighboring Mn spins, to two competing phases connected by a bi-critical point) and tunable quenched disorder. As the latter is increased, all the ordering temperatures decrease sharply in the vicinity of the bi-critical point, which is itself suppressed down to zero temperature, in analogy to the well-known quantum critical point (QCP) behavior. Upon a further increase of the quenched disorder, a charge-disordered paramagnetic (CDP) phase is stabilized in a wide $J_2$-T region, which includes a zero-temperature line. Short-range CO and FM correlations dominate this phase, often described as a "cluster glass". As supporting evidence to their computational work, Burgy *et al.* quote recent experimental data by Tomioka *et al.*[18]. The authors of these papers, by means of resistivity and magnetization measurements, identify regions of strong suppression of either FM or CE-type CO temperatures as a function of the tolerance factor, which can be attributed to QCP behavior. If the CDP is indeed present at all temperatures, it remains to be understood how it relates to the widely reported macroscopic phase separation, since the latter clearly implies long-range ferromagnetic and charge ordering. One possibility not to be discounted is that macroscopic phase separation may not be intrinsic, and that the "true" QCP behavior is only observed with particular combinations of cations or for exceptionally good-quality samples. An alternative explanation is that measurements of macroscopic properties are difficult to understand in a phase-separated regime, and a true picture can emerge only by combining them with microscopic probes of magnetic and charge ordering, such as neutron diffraction.

In order to clarify this important aspect of manganite phenomenology, we examined in detail, by means of neutron powder diffraction, magnetization and resistivity measurements, the part of the $Pr_{0.65}(Ca_ySr_{1-y})_{0.35}MnO_3$ phase diagram ($0.60 \leq y \leq 0.80$) purportedly containing a QCP for y=0.75[18]. This series was chosen not only because it is the same as in the work by Tomioka *et al.*, but also because dramatic changes in properties are achieved through a very narrow compositional range (y=0.8 is a CO insulator, y=0.6 a FM metal) with only minor changes of the *A*-site disorder. Our most significant finding is that *all* the ordering temperatures (the CO temperature $T_{co}$, the Neèl temperature $T_N$ and the Curie temperature $T_C$) are *continuous* through y=0.75. All of our data can be understood in terms of balance between long-range-ordered FM and CO phases, the only significant variable parameter being the relative phase fraction. The $T_{co}$ and $T_C$ curves cross at y=0.65, but mixed-phase samples are found on either side of the boundary. We find that the sharp upturn in the magnetization, which was identified by Tomioka *et al.* as a very low Curie temperature, corresponds in fact to a sudden increase in the amount of the ferromagnetic phase (phase fraction). This indicates that the CO phase is most stable in an intermediate temperature range, while the FM phase tends to dominate at low temperatures. Below 40 K, the AFM (pseudo-CE) structure associated with the CO phase undergoes a spin reorientation transition, possibly associated with simultaneous ordering of the Pr moments.

## 2. Experimental

Ceramic samples of $Pr_{0.65}(Ca_ySr_{1-y})_{0.35}MnO_3$ (y=0.60, 0.65, 0.7, 0.75, 0.80) were prepared by conventional solid-state reaction. Stoichiometric quantities of high-purity $Pr_6O_{11}$, $MnO_2$, and $ACO_3$ (A=Ca,Sr) were calcined at 900 °C and then sintered as pressed pellets for 24 hours at 1100 °C, 1200 °C, 1300 °C and 1350 °C in flowing oxygen with intermediate grinding. Following the final firing, the pellets were slowly cooled to room temperature. Resistance measurements were carried out with the conventional four-probe technique. Data were collected as a function of temperature on both cooling and warming at a rate of 2 K/min. Field-cooled DC magnetization data were collected on both cooling and warming at a rate of 2 K/min using a Quantum Design PPMS under a bias field of 0.5 T. Neutron powder diffraction data were collected using the GEM diffractometer at the ISIS neutron source (Rutherford Appleton Laboratory, UK). All measurements were performed on cooling from 280 K down to 10 K, using a closed-cycle refrigerator. The collection time was 15 min per temperature using samples of mass ~5 g, producing high-quality data in the d-spacing range $0.3Å≤d≤40Å$. Structural parameters, phase fractions, magnetic moments and spin orientations were refined with the Rietveld method using the program GSAS [19].

## 3. Neutron diffraction- General

The scattering function of all the samples is characterized by the simultaneous presence of nuclear and magnetic Bragg peaks (the latter at temperatures below $T_C$ or $T_N$ only), and by nuclear and magnetic diffuse scattering. Typical diffraction patterns, the contribution of each type of scattering being indicated, are shown in Figure 1. A long-range-ordering transition is signaled by the appearance of extra Bragg intensity, either coinciding with existing Bragg peaks (FM ordering) or at new positions of the reciprocal lattice (CO-AFM). The AFM and CO structures appearing throughout the whole compositional range are of the so-called pseudo-CE type, which has been described elsewhere[10, 14, 20]. In order for a Bragg peak to be observed, coherence of the crystal or magnetic structure must be established over domains of at least several hundred Ångstroms in size. Therefore, ordering temperatures are defined by the increase in intensity of corresponding Bragg peaks above their high-temperature values, and indicate long-range ordering over a part or all of the sample volume. Likewise, a spin reorientation transition is marked by the *relative* intensity changes of Bragg peaks belonging to the same magnetic structure, whilst preserving the overall value of the magnetic moment. The phase diagram of $Pr_{0.65}(Ca_ySr_{1-y})_{0.35}MnO_3$, constructed by drawing phase lines through the onset ordering temperatures as defined above, is shown in Figure 2.

We find that all of the ordering temperatures are continuous through the y = 0.75 composition. However, this is in itself insufficient to disprove the QCP scenario, as ordering could involve only a small part of the sample, perhaps with a slightly different composition. We will therefore investigate the possible existence of the CDP phase, a key ingredient of the QCP model. Our argument will proceed as follows: 1) We will show that, whenever $T_{co} > T_C$ (y≥0.70), the whole sample undergoes a transition to a well-ordered CO phase at $T_{co}$. 2) At $T_C$, part of the sample becomes FM. At this point, the CO and FM phases coexist. 3) Below $T_N$, the CO phase becomes AFM. At this point, the whole sample is magnetically ordered for all compositions. 4) In this series of samples, there is no trace at low temperatures of short-range ferromagnetic fluctuations, which are the hallmark of the CDP phase both in theory and experiments, and are present, as expected, at high temperatures in the "true" paramagnetic phase. We conclude that the low-temperature CDP phase, if it exists at all, is only present as a small fraction. In these samples, the metal-to-insulator transition is dominated by percolation in a macroscopic CO-FM phase assemblage.

## 4. Single-Phase Refinements

Of all the compositions we studied, the y=0.80 is by far the easiest to understand, since neutron diffraction patterns show no trace of ferromagnetic behavior. This is in agreement with the magnetization measurements discussed later: even if the small ordered moment (0.25 $\mu_B$/Mn) were preserved in zero field, it would be too small to be observed by neutron powder diffraction. Since y=0.80 is well to the Ca-rich side of the "QCP", this sample is expected to display CO with a high degree of ordering, and to be free of the CDP phase. This scenario is in excellent agreement with the results of our single-phase refinements, based on the CO model we have previously proposed[14]. In fact, the quality of the present data is such that we can freely refine *all* the structural parameters of the CO phase in the $P2_1/m$ space group. The only constraints used were on the isotropic Debye-Waller factors; these were constrained to be equal on the two $Mn^{3+}$ sites, on the four Pr/Ca/Sr sites, for the "in-plane" and for the "out-of-plane" oxygen atoms. Figure 3 shows the evolution of the Jahn-Teller distortion parameter ($\sigma_{JT}$) with temperature, averaged over the two $Mn^{3+}$ sites. This parameter is defined as

$$\sigma_{JT} = \left[ 1/3\sum_i\left((Mn\text{-}O)_i - <Mn\text{-}O>\right)^2 \right]^{1/2}$$

where $(Mn\text{-}O)_i$ is the $i^{th}$ bond distance in an $MnO_6$ octahedron and $<Mn\text{-}O>$ is the mean bond distance [21]. $\sigma_{JT}$ is a measure of the degree of orbital ordering in the CO phase. The low-temperature value of $\sigma_{JT}$=0.062 Å is similar to that of the CE-CO phase in La$_{0.5}$Ca$_{0.5}$MnO$_3$ ($\sigma_{JT}$=0.068 Å) [22], but small in comparison with purely Mn$^{3+}$ manganites such as LaMnO$_3$ ($\sigma_{JT}$=0.116 Å) [23], PrMnO$_3$ ($\sigma_{JT}$=0.133 Å) [24] and HoMnO$_3$ ($\sigma_{JT}$=0.134 Å) [25]. The magnitude and direction of the ordered AFM moments on both the Mn$^{3+}$ and Mn$^{4+}$ sublattices were also refined, and their temperature dependence was found to be consistent with a usual power-law behavior. The moments always lie in the xy plane; for T>40 K they are at an angle of ~20$^\circ$ to the b-axis on the Mn$^{3+}$ sublattice and essentially parallel to the b-axis on the Mn$^{4+}$ sublattice. Additional confirmation of full ordering is provided by the very large refined AFM moment at low temperatures (3.05(1) $\mu_B$/Mn averaged over both sublattices). This is larger than the corresponding ordered moment found in La$_{0.5}$Ca$_{0.5}$MnO$_3$ (2.78(1) $\mu_B$/Mn)[22], although still smaller than the single-ion value for this doping (3.65 $\mu_B$/Mn). In addition to covalency effects, this reduction is most likely due to the presence of magnetic defects, as found in La$_{0.5}$Ca$_{0.5}$MnO$_3$ [22]

All the other samples display FM behavior at low temperatures. However, for (y≥0.70), where $T_C < T_{co}$, there exists a temperature region $T_C < T < T_{co}$ where the patterns can be satisfactorily analyzed in the same way as for y=0.8, i.e., with a single-phase CO model. The refined values of $\sigma_{JT}$ for these T-y points are also shown in Figure 3. If CO were giving way to CDP behavior, we would expect orbital ordering to become increasingly weak on approaching the QCP composition. On the contrary, Figure 3 clearly shows that all three compositions are equally well ordered above $T_C$. The y=0.60 patterns at T < 50 K contain a single FM phase, with no trace of CO. Once again, the high value of the refined moment (3.45(3) $\mu_B$/Mn, aligned parallel to the a-axis; here the slight reduction with respect to the single-ion value is probably due to covalency) rules out the presence of a CDP component.

## 5. Refinements in the Two-Phase Region

A large number of T-y points contain FM, CO and AFM peaks, which, in these compounds, is indicative of phase coexistence[26]. Since the resolution of GEM ($\Delta d/d \geq 0.25\%$) does not allow nuclear peaks of the coexisting phases to be resolved, unconstrained multi-phase refinements are not viable, and one needs to implement strategies to extract the maximum amount of relevant information. We have chosen two basic approaches.

Firstly, refinements can be performed with a *single phase model*, allowing the FM and AFM moments to vary independently. This strategy provides satisfactory fits, because the difference in lattice parameters between the two phases is small compared to the instrument resolution. The sum in *quadrature* of the refined FM and AFM moments yields the root mean squared total ordered moment per manganese site, $M_{RMS}$, which is plotted in Figure 4; this is the only magnetic parameter that can be obtained without any knowledge of the phase fractions of the coexisting phases[27]. In this case, the nuclear phase was chosen to have *Pnma* symmetry, and no information regarding the superstructure was gathered. The QCP scenario has a very strong implicit prediction regarding the behavior of $M_{RMS}$: this parameter should be strongly suppressed as the CDP phase sets in, because disordered spins do not contribute to the intensity of Bragg reflections. Figure 4 clearly demonstrates that our samples do not adhere to this prediction. At low temperatures, $M_{RMS}$ is always greater than the full AFM ordered moment (~3.1 $\mu_B$/Mn for y=0.80), and increases monotonically with decreasing y up to the full FM ordered moment (~3.5 $\mu_B$/Mn for y=0.60). As already anticipated, there is no trace of a magnetically disordered phase for y=0.75. The slope changes corresponding to lower-temperature transitions are also noteworthy. These would not be expected for spin reorientation/canting, and confirm indirectly the presence of multiple magnetic phases.

A reorientation of the AFM moments from the direction of the b-axis towards the a-axis was detected for $0.70 \leq y \leq 0.80$ at or below 40 K (see Figure 2). This transition, which is most likely associated with ordering of the Pr moments, was not previously reported for the pseudo-CE phase, and is described in more detail in Appendix I. Clear signatures of the metallization and CO transitions are visible in the lattice parameters and atomic displacement parameters of the *Pnma* crystallographic phase. These anomalies are not central to the present discussion, but can serve as guides to the underlying electronic phenomena within a simple structural model. We have reported them in more detail in Appendix II.

As an alternative approach, constrained 2-phase refinements (a CO-AFM phase with space group $P2_1/m$ in combination with a FM phase with space group *Pnma*) can be performed. Here, the principal aim is to obtain the phase fractions of the coexisting phases. Therefore, we have fixed both structural and magnetic parameters of the CO-AFM phase to be equal to those of the y=0.80 sample at each corresponding temperature. As we have already seen, this approach is supported by the refinements in the single-phase region, since $T_N$, $\sigma_{JT}$ and the AFM moment are essentially composition-independent. Also, in the mixed-phase regions we have verified that the intensities of the AFM and CO peaks are always linearly correlated (this is in itself a very strong indication of phase segregation as opposed to spin canting). The second phase was chosen to have *Pnma* crystallographic symmetry with a refineable magnetic moment[28]. Figure 5 shows the refined fraction of the *Pnma*-FM phase throughout the phase diagram. The y=0.70 and y=0.75 samples display the same sequence of transitions: at $T_{CO}$, long-range, CO rapidly sets in over nearly all the sample volume. At $T_C$, a small portion of the sample acquires a FM moment. At lower temperatures (40-100 K), the FM-*Pnma* phase fraction shows a further increase, followed by a final saturation at low temperatures. The y=0.65 sample shows a similar increase in the FM-*Pnma* phase fraction between 60 and 110 K; although $T_{CO}$ and $T_C$ coincide, the CO phase initially dominates. The sample-dependent quantities in Figure 5 are the values of the phase fraction at the high- and low-temperature "plateaus" and the temperature $T_{C2}$ where the transition between the two occurs. We have defined $T_{C2}$ as the mid-point of the phase fraction increase between the two plateaus. As can be seen in Figure 2, $T_{C2}$ has a similar composition dependence to the values of $T_C$ as determined by Tomioka *et al.*[18], but has clearly a different meaning: $T_{C2}$ is not a magnetic ordering temperature, but a point where the balance of phases is dramatically altered. The y=0.60 sample, where $T_C>T_{CO}$, displays a slightly different behavior within the same overall scenario. Clearly, a large FM-*Pnma* fraction is established *before* long-range CO sets in. In Figure 5 this fraction is shown as 100%, but it is likely that part of the sample is still in the high-temperature disordered state; the present data do not allow the true fraction to be determined. On cooling CO briefly competes, reaching a maximum of about 30% phase fraction, but is completely suppressed at low temperatures.

Figure 6 shows the FM moments of the *individual* phases, where refined: clearly, these parameters are rather insensitive to the composition. This is further confirmation of the fact that most of the competition between different ordering schemes occurs on the lengthscale of hundreds of angstroms, not on the nanoscopic scale, and that the phase segregation model is correct. In the QCP scenario, we would have expected the *Pnma* phase to be partially paramagnetic with a much-reduced magnetic moment.

6. Short-range structural and magnetic ordering

One of the peculiarities of the proposed CDP phase is the fact that it contains CO and FM domains of nanometer size, i.e., much smaller than the correlation length of the neutron diffraction probe. The CDP phase should therefore be associated with enhanced nuclear and magnetic diffuse scattering. Both features can be probed in our data.

In order to investigate the short-range structural order, we calculated the pair-distribution functions (PDF) for all our data sets. The PDF is a representation of the local structure in real space and makes use of both Bragg peaks and diffuse scattering. This technique has been widely used to describe local disorder in crystalline materials, including manganites[29]. The PDF's were extracted using the merged GEM histograms between 0.9 Å$^{-1}$ and 28 Å$^{-1}$. While the higher Q limit was imposed by the high-Q statistics of the data, the lower Q limit was chosen in order to minimize the magnetic contribution to the PDF while still including the nuclear Bragg peaks. In this way the broad oscillations characteristic of the magnetic contribution to the PDF are negligible in the data in both FM and AFM regions. The experimental PDF's were fitted with a very simple 2-phase model, which describes the data as a mixture of a localized, Jahn-Teller distorted phase and a "delocalized" (D) phase with regular octahedra and no Jahn-Teller distortion. Only data up to 4.1 Å were considered, making our model sensitive only to short-range order. The models for the two "pure" phases were taken from the y=0.80 and y=0.60 low-temperature data sets, respectively, and allowance was made for thermal broadening.  A single parameter, the phase fraction of the D phase, was extracted from these fits.  It is noteworthy that this model cannot distinguish between the paramagnetic and the CO phase, because both support JT polarons (long-range ordered in the latter).  The phase fraction of the D phase is plotted in Figure 5 together with the Rietveld FM phase fraction as a function of temperature and composition.  There is excellent agreement between the two parameters, which, we stress, are determined in a completely independent way.  The only significant discrepancy occurs at high temperatures for the y=0.60 FM sample, and reflects the well-known presence of JT polarons well into the FM phase even for metallic samples (see for example Reference 29).  In the QCP scenario, we would have expected a significant discrepancy near y=0.70:  in fact, the CDP phase would contribute to the *Pnma* phase in the Rietveld analysis and to the JT phase in the PDF analysis.

The magnetic diffuse scattering was extracted from the data by subtracting the Rietveld-refined Bragg peaks from the data sets at low momentum transfer Q.  The presence of oscillations, which are characteristic of short-range magnetic order, are clearly visible in the paramagnetic region.  Before Rietveld refinement, the data were normalized to the scattering intensity obtained from a vanadium rod.  For the diffuse-scattering analysis, the Bragg pattern-subtracted data were then replotted in absolute units (barns/sterad/formula unit) by use of the Rietveld-refined scale factor, and the nuclear incoherent neutron scattering cross section was subtracted.  For all compositions, the room-temperature scattering intensity increases rapidly below $Q{\approx}0.5$ Å$^{-1}$, a characteristic signal of short-range FM ordering.  In the paramagnetic region, the data below Q=0.5 Å$^{-1}$ could be fitted well using the spin correlation model of Viret *et al.* [30], which yields the short-range magnetic correlation length and the associated magnetic moment per formula unit. The fits were extended to include diffuse scattering in the region of the FM Bragg peak at Q=1.64 $Å^{-1}$, approximated by a Gaussian line shape. A typical fit (for $y$=0.75, T=280 K) is shown in the inset to Figure 7. The FM correlation length for all compositions increases slightly, from 3.8(5) Å to 4.5(5) Å on approaching $T_C$. This parameter does not diverge at $T_C$ because of the well-known "central peak" behavior[31]. The associated magnetic moments, at 2.4(2) $\mu_B$/Mn, show no obvious variation with temperature below 280 K and are essentially sample-independent. For y≥0.65 additional diffuse scattering near the positions of the AFM Bragg peaks becomes apparent at temperatures ranging from T<250 K for y=0.8 to T<190 K for y=0.65. The low-Q FM diffuse intensity rapidly decreased on cooling through $T_C$ and $T_N$. At low temperatures, the statistical quality and the Q-range of our data are not sufficient to estimate the magnetic moments and correlation lengths independently. Therefore, we illustrate the systematic behavior of the short-range magnetic correlations by plotting, in Figure 7, the average value of the magnetic scattering cross section at Q=0.4 $Å^{-1}$ (the integration window was 0.2 $Å^{-1}$). All compositions display similar behavior, with the diffuse intensity rapidly decreasing on cooling through $T_C$ or $T_N$. The different behavior of the curves reflects the different shape of the FM and AFM magnetization. Once again, the diffuse magnetic scattering supports phase segregation between two long-range-ordered magnetic phases, and not the presence of a "paramagnetic-like" phase down to low temperatures.

## 7. Magnetization and transport measurements

The nature of the competition between the FM and CO-AFM phases at the mesoscopic/macroscopic level is well illustrated by our magnetization and transport measurements, which agree well with previously published data[18], but are also completely consistent with our neutron diffraction results. Plots of magnetization versus temperature measured on heating and on cooling in an applied field of 0.5 T are shown in Figure 8. Considering first the cooling curves, the y = 0.6 and 0.65 samples display a sudden upturn in the magnetization below ~210 K, corresponding to the onset of long-range FM ordering. In both samples, the magnetization approaches a value of ~3.2 μB/Mn at low temperatures. For y = 0.65 a significant proportion of the sample is in the CO-AFM state on cooling through 200 K, and a second sudden upturn in the magnetization, evident at ~90 K, may be attributed to the sudden increase in the FM phase fraction at $T_{C2}$ observed by neutron diffraction, approaching 100% at 10 K. For the y = 0.7, 0.75 and 0.8 samples no obvious onset of long-range FM ordering is apparent. Rather, with increasing y, the shape of the magnetization curve becomes more typical of a transition to long-range AFM ordering close to 200 K. There are upturns in the magnetization on cooling through ~50 K for the y = 0.7 and 0.75 samples, which again correspond to increases in the FM phase fraction at $T_{C2}$; however, the magnetization at low temperatures is much smaller than that of the y = 0.6 and 0.65 samples, since the FM phase fraction never reaches 100%. The low-temperature magnetization is only ~0.25 μB/Mn for y = 0.8, where the CO-AFM phase dominates over the whole temperature range. The slight increase in magnetization on cooling is most likely due to a partial ferromagnetic ordering of the Pr moments.

The large divergence of the magnetization curves measured on cooling and heating for y = 0.65 to 0.8 illustrates the hysteretic FM/AFM competition in the phase-separated region of the phase diagram. Initially, the CO-AFM phase becomes well established at near 200 K, creating a strain field that the FM phase, becoming the stable phase as the temperature is lowered, must overcome. Conversely, the strain field of the FM-phase suppresses the CO-AFM phase on heating. For samples in the phase-separated region, there will be a large hysteresis between the relative phase fractions on heating and cooling through any given temperature. This picture is supported by measurements of electrical resistance versus temperature, shown in Figure 9. Considering first the cooling curves, insulator to metal transitions are apparent for the y = 0.6 to 0.75 samples. These transitions occur at $T_C$ for y = 0.6 and at progressively lower temperatures as y increases, corresponding well to $T_{C2}$ in the neutron diffraction measurements and to the sudden upturns in magnetization. The low-temperature resistance increases as the CO-AFM phase becomes more dominant with increasing y. Indeed, the y = 0.8 sample is insulating throughout the temperature range measured. In analogous fashion to the magnetization, the FM/AFM competition at the macroscopic level gives rise to a hysteresis of the resistance curves for samples in the two-phase region.

## 8. Discussion

Our data do not support the QCP scenario as a framework for the series of transitions occurring in the $Pr_{0.65}(Ca_ySr_{1-y})_{0.35}MnO_3$ phase diagram. The crucial ingredient, i.e. the low-temperature CDP phase, is notably absent in the present case. It is noteworthy that a charge-disordered insulating phase, sometimes referred to as an orbital glass, does in fact exist elsewhere in the phase diagram, for example in $Pr_{0.7}Ca_{0.3}MnO_3$ [20]. This phase is weakly ferromagnetic with a reduced magnetic moment and low $T_C$, and is probably related to the ferromagnetic insulating phase, with reduced spin stiffness constant, that is ubiquitous at low doping [32, 33]. At the present doping level of 0.35, and with this combination of cations, the CDP phase does not form.

Instead, the $Pr_{0.65}(Ca_ySr_{1-y})_{0.35}MnO_3$ phase diagram is dominated by mesoscopic/macroscopic phase coexistence. The theoretical framework for phase coexistence in these materials is now well established. Computational studies have shown that magnetic exchange disorder on an atomic scale (due, for example, to compositional fluctuations) can stabilize "giant" clusters on a much longer lengthscale[34]. Also, the relative stability of the coexisting phases is strongly influenced by intra-granular stresses, similar to martensitic phase transitions[16, 20, 35]. As expected, the relative proportions of the coexisting phases change as a function of y (Figure 2). Much less expected is the striking temperature dependence of the phase fractions, with the FM phase being rapidly stabilized below a characteristic temperature $T_{C2}$. This strongly suggests the presence of a temperature-dependent CO-FM phase line in the "ideal" phase diagram (i.e., in the absence of quenched disorder and strain). It is unclear why the CO phase should be more stable at higher temperatures; we speculate that this might be related to its increased entropy, due to the presence of charge/orbital disorder on the nominal $Mn^{4+}$ sites, which have to accommodate 30% of $Mn^{+3}$. In any case, reproducing this picture will represent a significant challenge for the computational and theoretical modeling of these materials.

## 9. Appendix I: Spin reorientation and Pr moment ordering at low temperatures

Partial ordering of the Pr moments occurs at low temperatures. The onset temperature is impossible to define given the gradual nature of the ordering, but extra intensity in the ferromagnetic peaks begins to appear below ~100 K for the y = 0.6 sample. The directions of the FM-ordered Pr moments could not be refined, and they were constrained to lie along the a-axis. The ordered Pr moment at 10 K in the y = 0.6 sample is 0.40(2) $\mu_B$/Pr; it decreases with increasing y and becomes too small to refine accurately in the y = 0.8 sample.

In the AFM phase of the 0.70≤y≤0.80 samples a spin reorientation occurs at ~40 K. Changes in the relative intensities of the (101) and (100) AFM peaks indicate a reorientation of the spins from close to the b-axis towards the a-axis; in the xy plane the spins on both the $Mn^{3+}$ and $Mn^{4+}$ sublattices rotate by ~50° between 20 K and 40 K. No anomalies in the crystal structure or in the Debye-Waller factors are apparent in this temperature interval; the cause of the spin reorientation is at present unclear, but one possibility is that it may be induced by the partial ordering of Pr moments.

## 10. Appendix II: Lattice and ADP anomalies at the ordering temperatures

The temperature dependence of the lattice parameters and unit cell volume, as obtained from single-phase Rietveld refinements with *Pnma* crystallography symmetry, is plotted in Figure 10. For the y = 0.6 sample a clear decrease in the *a* lattice parameter and in the cell volume occurs at $T_C$ = 210 K. This is most likely due to the suppression of incoherent JT distortions as long-range FM ordering sets in. There are obvious anomalies in the lattice parameters at $T_{CO}$ for the 0.65≤y≤0.80 samples. The onset of charge and orbital ordering on cooling, with the corresponding increase of coherent JT distortion, causes a decrease in *a* and *b* and a large increase in *c*, the cell volume smoothly decreasing with temperature.

The temperature dependence of the anisotropic Debye-Waller factors for the "apical" oxygen atoms O1 and the "in-plane" oxygen atoms O2, as obtained from single-phase

Rietveld refinements with *Pnma* crystallography symmetry, is plotted in Figure 11. The values shown are the projections of the thermal ellipsoids both parallel and perpendicular to the Mn-O bonds. For y = 0.6 and 0.65, there is a sudden decrease in all the Debye-Waller factors as the sample is cooled through $T_C$. This is indicative of the suppression of incoherent JT distortions as the insulator-metal transition is crossed, as reported elsewhere [29, 36, 37]. More interestingly, we observe clear signatures of the CO transition, with a sudden decrease of the *longitudinal* component for the apical oxygen O1 and an increase of the *transverse* component for the in-plane oxygen O2. Both observations are consistent with a rotation of the filled $e_g$ orbitals in the a-c plane and the development of a pattern of in-plane oxygen displacements.

## 11. Acknowledgements

example Reference 20). For these data, additional confirmation is provided by the orientation of the spins (both FM and AFM moments have a significant a-axis component), the temperature dependence of the individual FM and AFM components and the scaling of the AFM and CO reflections (see below).

# Figure Captions

<u>Figure 1</u>: Typical GEM neutron diffraction patterns for $Pr_{0.65}(Ca_ySr_{1-y})_{0.35}MnO_3$ (the plots shown are for $y = 0.65$ at 100 K). **Main panel:** observed, calculated and difference diffraction profiles for a Rietveld refinement of nuclear and magnetic structures. The refined data were collected from four detector banks situated at $2\theta =$ 17.9º, 63.5º, 91.4º and 153.9º, and refined simultaneously. To produce the figure, data from different banks in adjacent d-spacing ranges were spliced at points of the profile where no Bragg peaks are present. Tick marks show the peak positions for (from top to bottom) the FM magnetic phase, the *Pnma* nuclear phase, the CO nuclear phase, and the AFM magnetic phase. Superlattice peaks from the CO phase (**left inset**, marked with arrows) are evident, as is extra intensity in some nuclear peaks due to FM ordering (**right inset**), where a dashed line indicates the contribution from the nuclear phase.

<u>Figure 2</u>: Schematic phase diagram of $Pr_{0.65}(Ca_ySr_{1-y})_{0.35}MnO_3$. $T_{CO}$, $T_N$ and $T_C$ are the onset temperatures of charge ordering, AFM ordering and FM ordering, respectively. A sudden increase in the FM phase fraction and electrical conductivity occurs at $T_{C2}$ (see text for details) and a spin reorientation in the AFM phase occurs at $T_{SR}$. The majority phase (i.e., with phase fraction > 50%) is indicated by different shades of gray: paramagnetic insulator (white), ferromagnetic metal (light gray) and charge-ordered insulator (dark gray). The hatched area indicates phase coexistence.

<u>Figure 3</u>: Temperature dependence of the Jahn-Teller distortion parameter $\sigma_{JT}$ of the CO-phase $Mn^{3+}$ site. Values are plotted only for the single-phase regions of the y=0.70, 0.75 and y=0.80 compositions.

<u>Figure 4</u>: Temperature dependence of the root mean squared total ordered magnetic moment per Mn atom. Error bars are smaller than the symbols.

<u>Figure 5</u>: Temperature dependence of the ferromagnetic/charge-delocalized phase fraction for $0.6 \leq y \leq 0.75$. **Closed symbols** are for the FM- *Pnma* phase from 2-phase Rietveld refinements (error bars are approximately ±0.005). **Open symbols** are for the delocalized (non-Jahn-Teller distorted) component in the PDF analysis (see text; error bars are approximately ±0.07). For each composition, the approximate position of $T_{C2}$, where there is a pronounced increase of the FM phase fraction, is also indicated (see text).

<u>Figure 6</u>: Temperature dependence of the FM ordered moment for $0.6 \leq y \leq 0.75$. The line connecting the y=0.6 data is a guide to the eye.

<u>Figure 7</u> **Main panel:** Average value of the magnetic diffuse scattering intensity in the range $0.3 \leq Q \leq 0.5$ Å$^{-1}$ as a function of temperature. **Inset:** Typical raw neutron diffraction data after subtraction of Bragg peaks and the incoherent neutron scattering. The plot shown is for $y = 0.75$, $T = 280$ K. The solid line is a fit obtained by adjusting the magnetic moment and correlation length (see text).

<u>Figure 8</u>: Field-cooled magnetization versus temperature for $0.6 \leq y \leq 0.8$. Data were collected both on cooling and heating (indicated with arrows) in an applied magnetic field of 0.5 T.

Figure 9: : Electrical resistance normalized to the $T$=295 K value versus temperature for $0.6 \leq y \leq 0.8$. Data were collected both on cooling and heating (as indicated by arrows).

Figure 10: Lattice parameters and unit-cell volume as a function of temperature, obtained by fitting the data with a single *Pnma* nuclear phase (see text). Error bars are smaller than the symbols.

Figure 11: Projections of the oxygen anisotropic Debye-Waller factors *parallel* ($U_{par}$) and *perpendicular* ($U_{perp}$) to the direction of the Mn-O bond lengths for the O1 (apical) and O2 (in-plane) atoms. Plots are shown for three representative compositions: y=0.6 (top), 0.7(middle) and 0.8 (lower). These values were obtained by fitting the data with a single *Pnma* nuclear phase (see text).

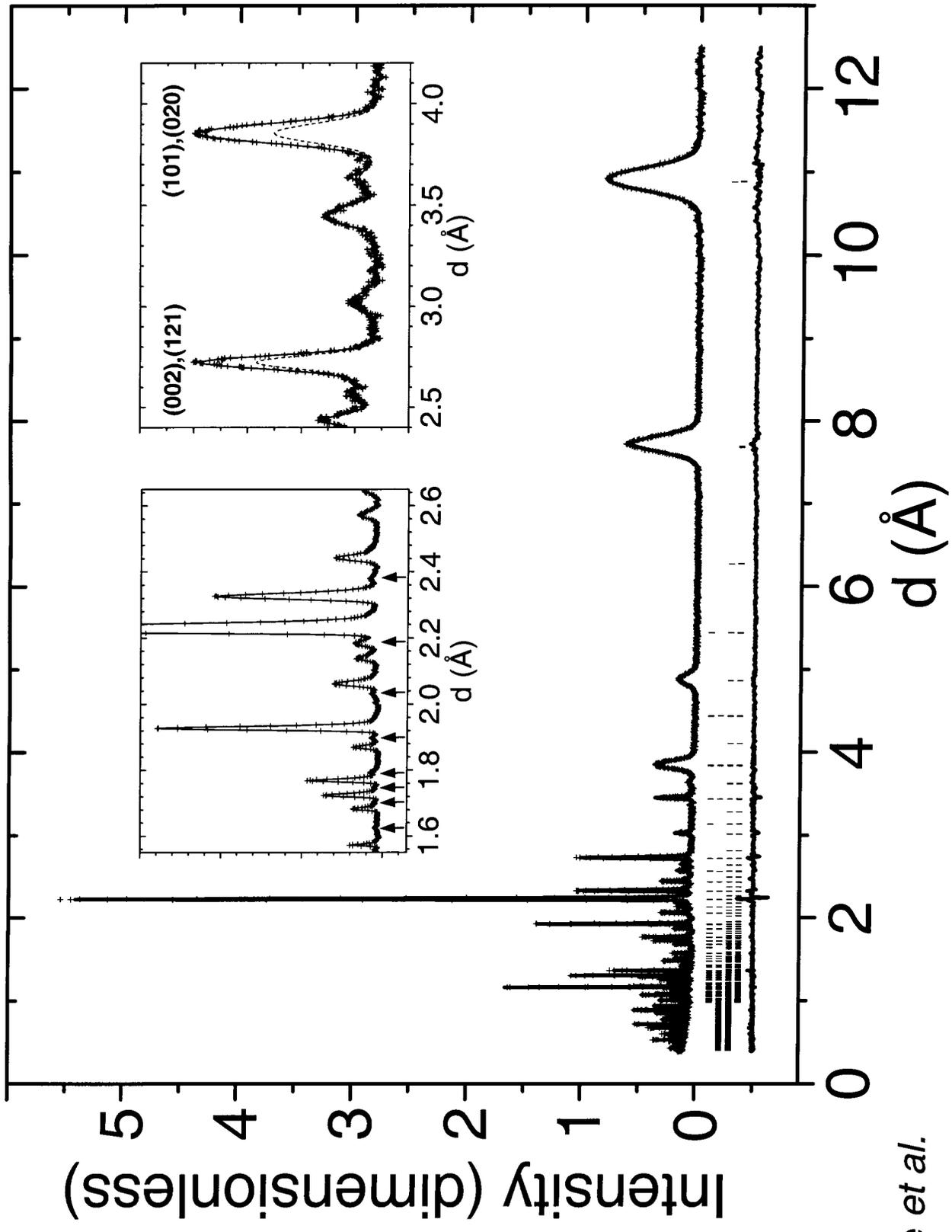

G. Blake et al.
Figure 1

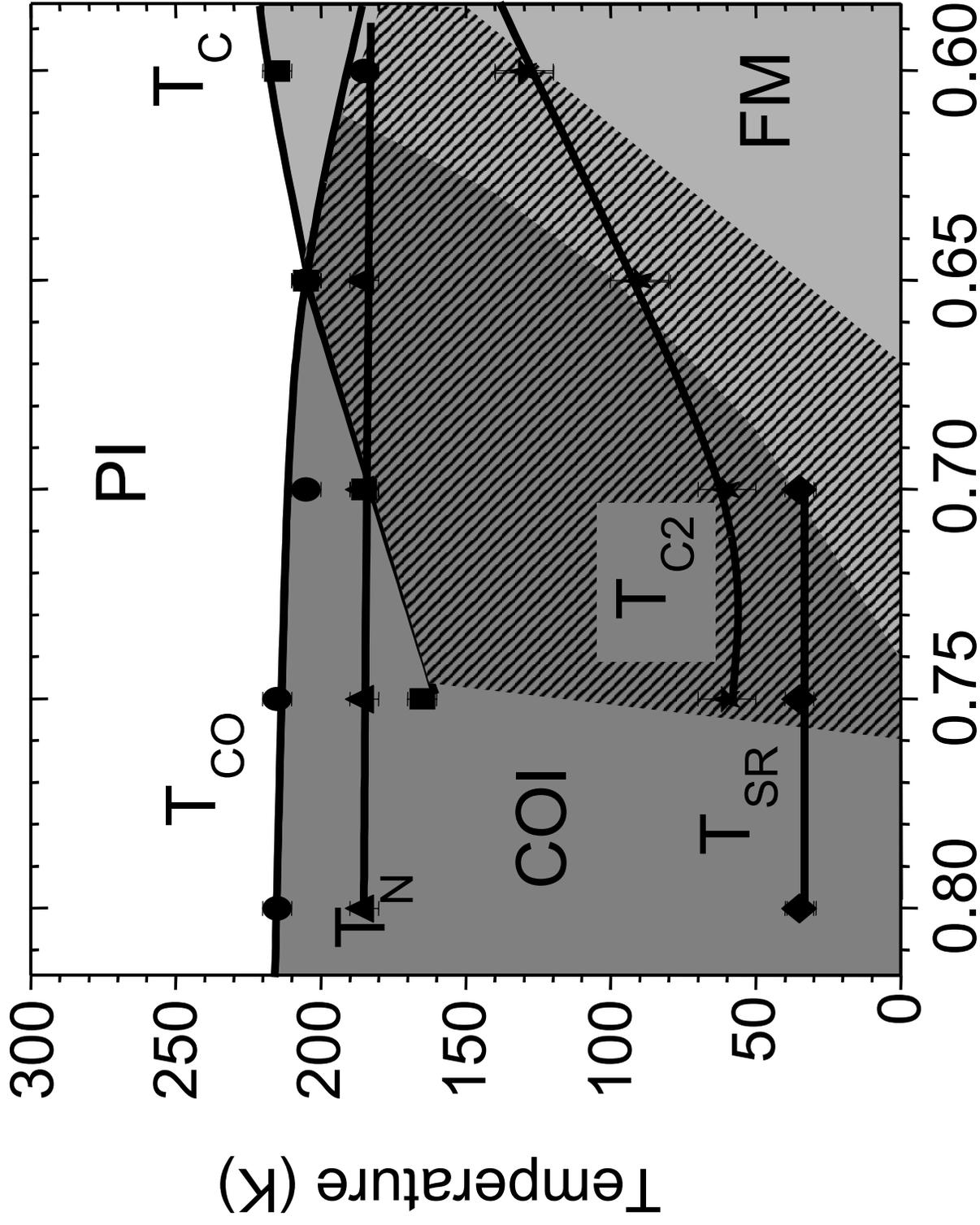

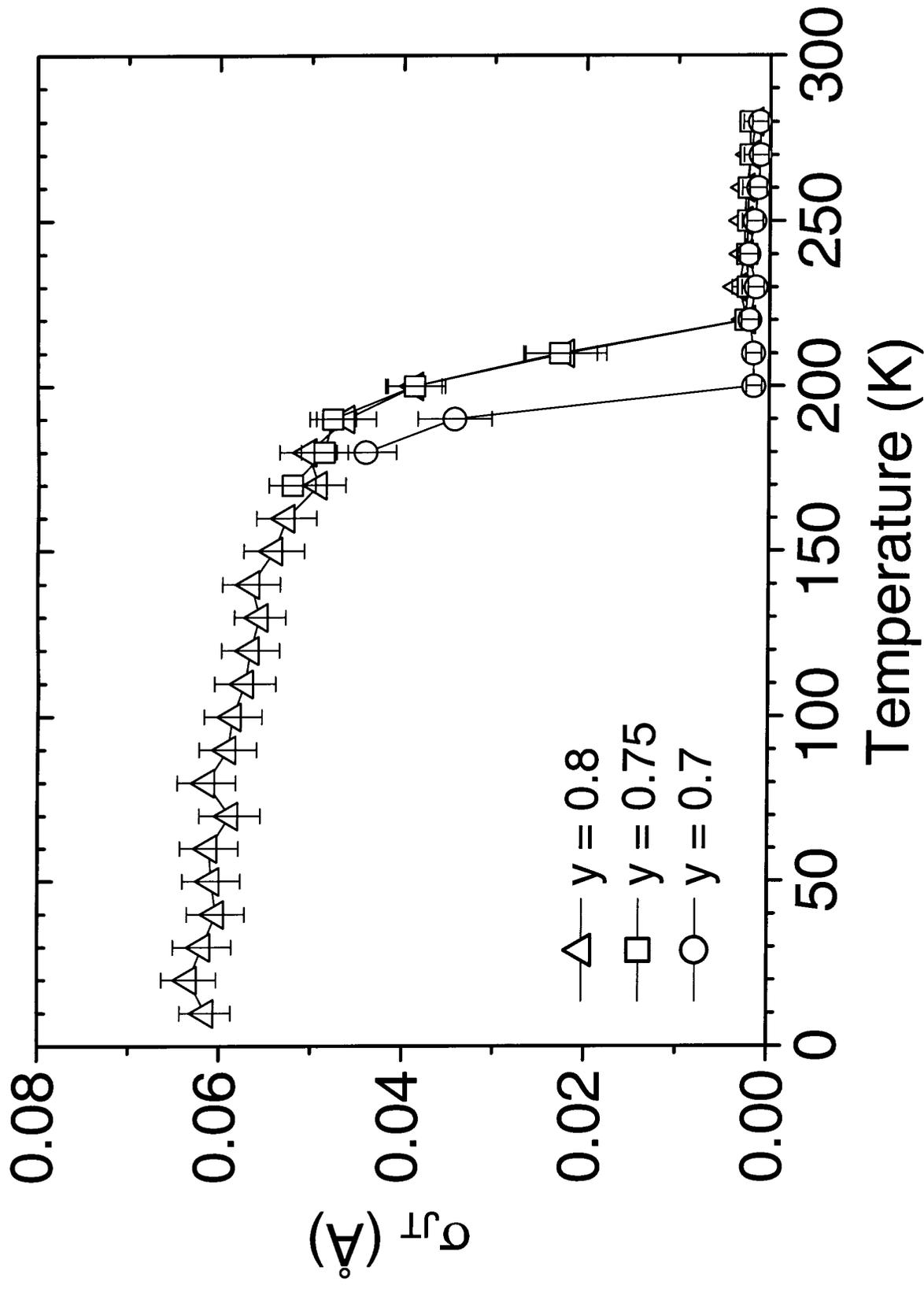

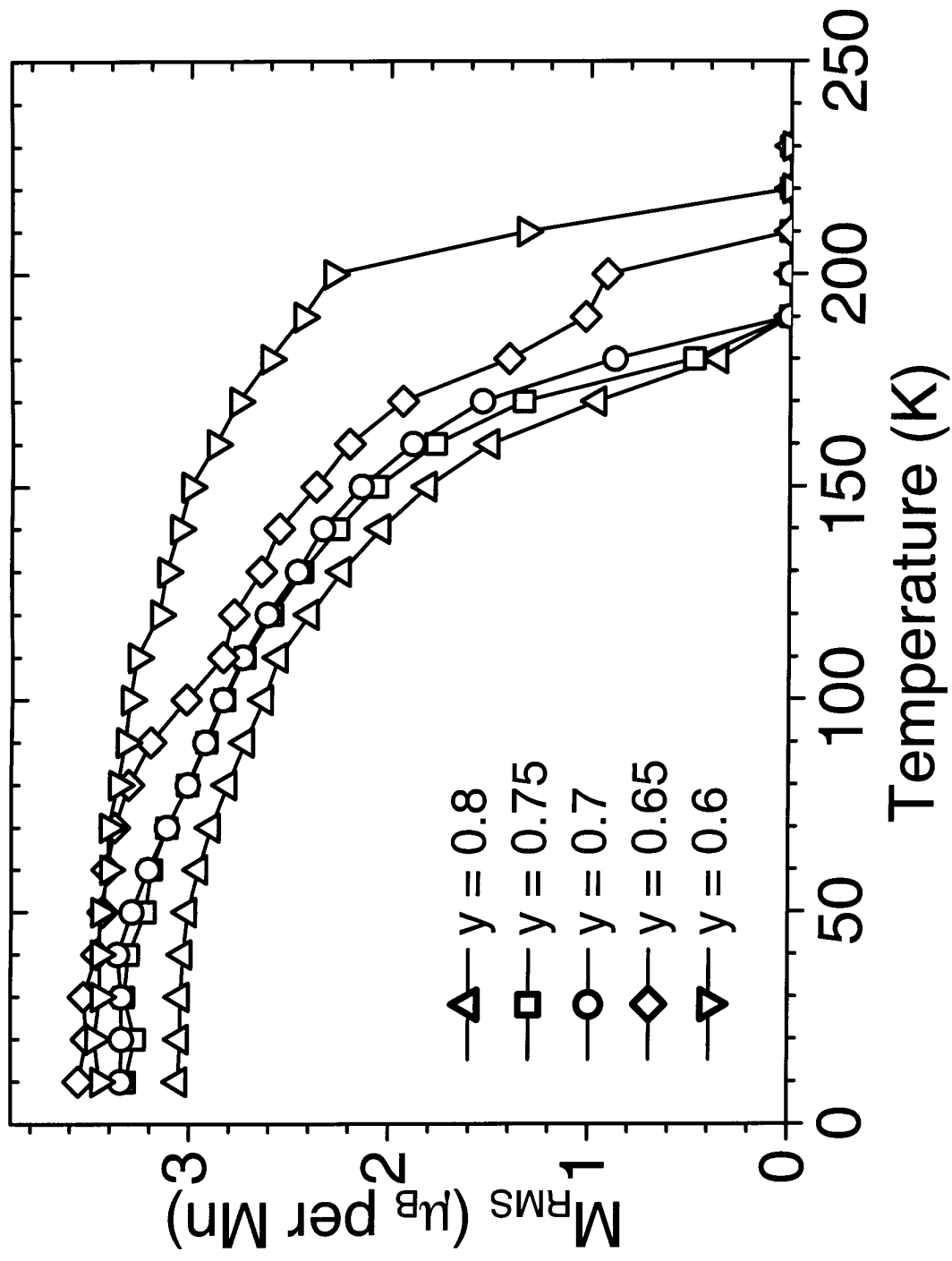

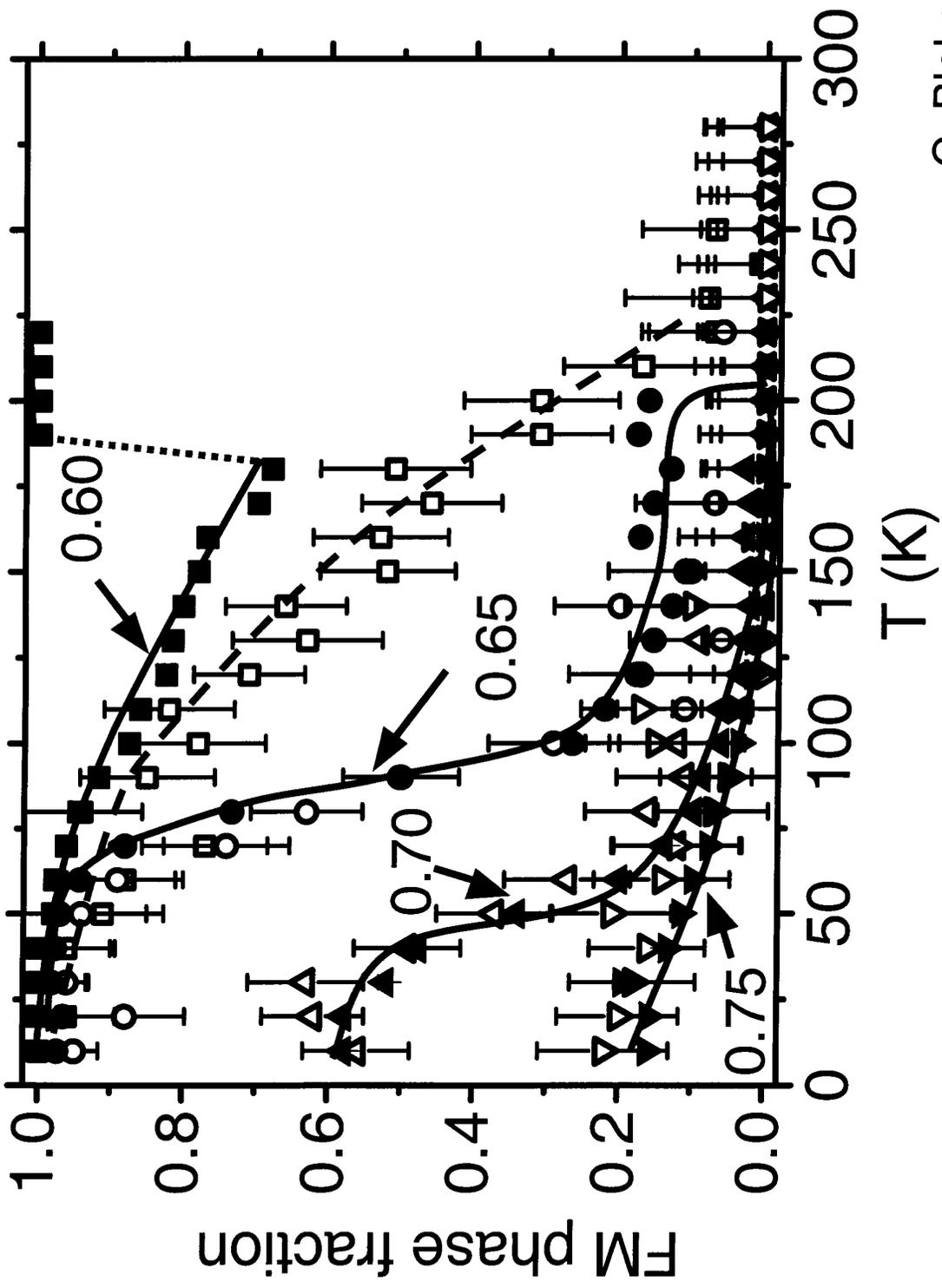

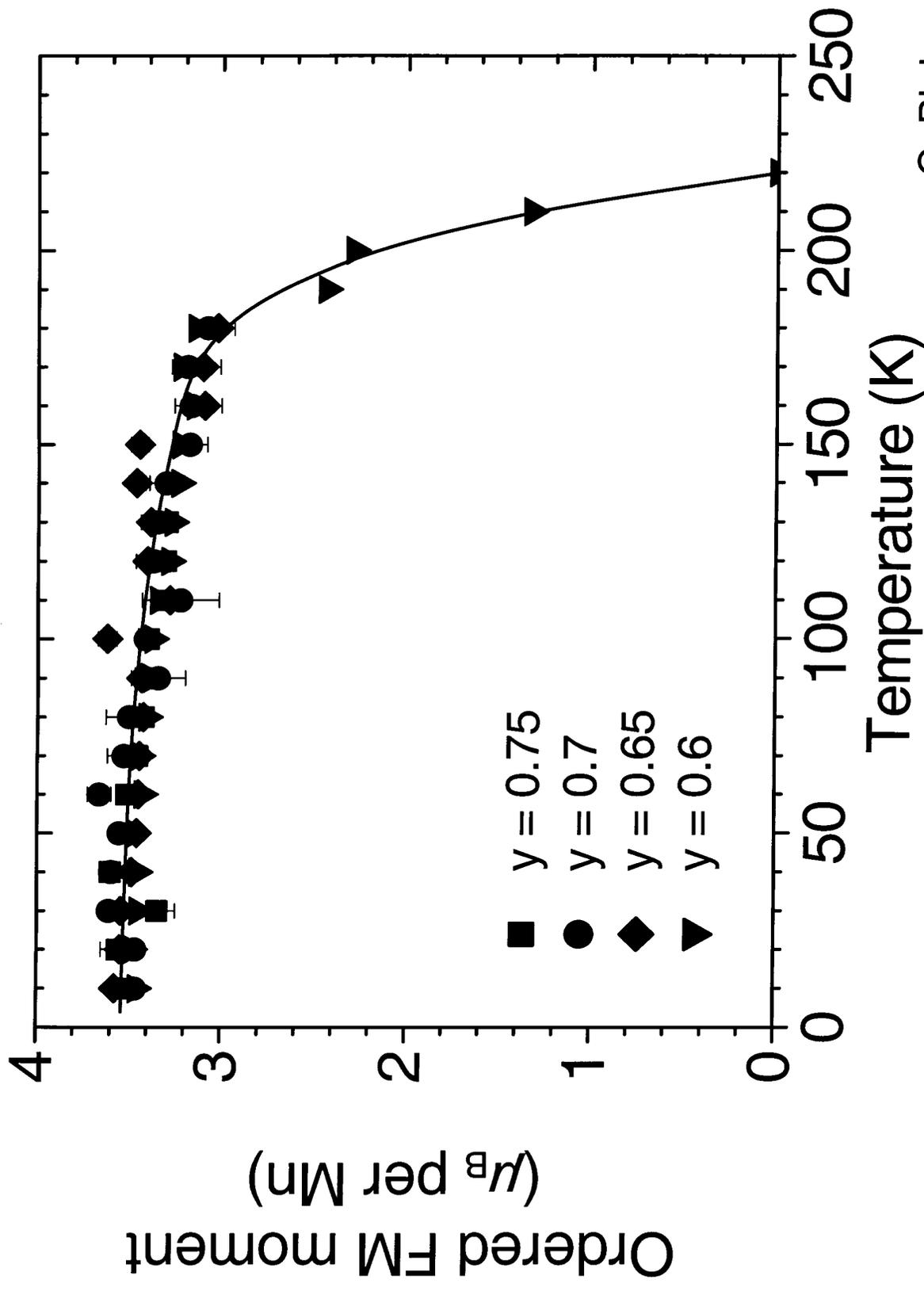

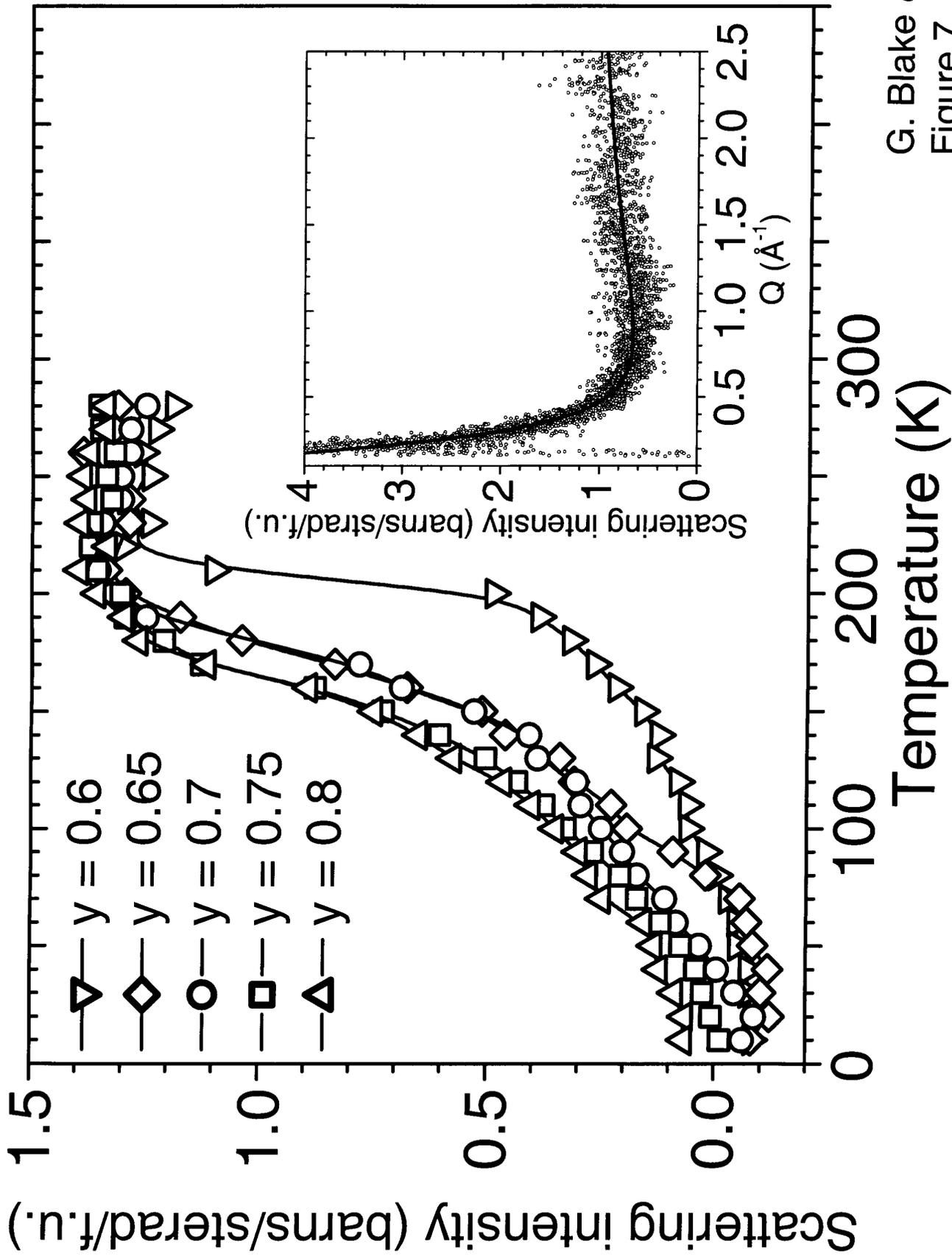

G. Blake *et al.*
Figure 7

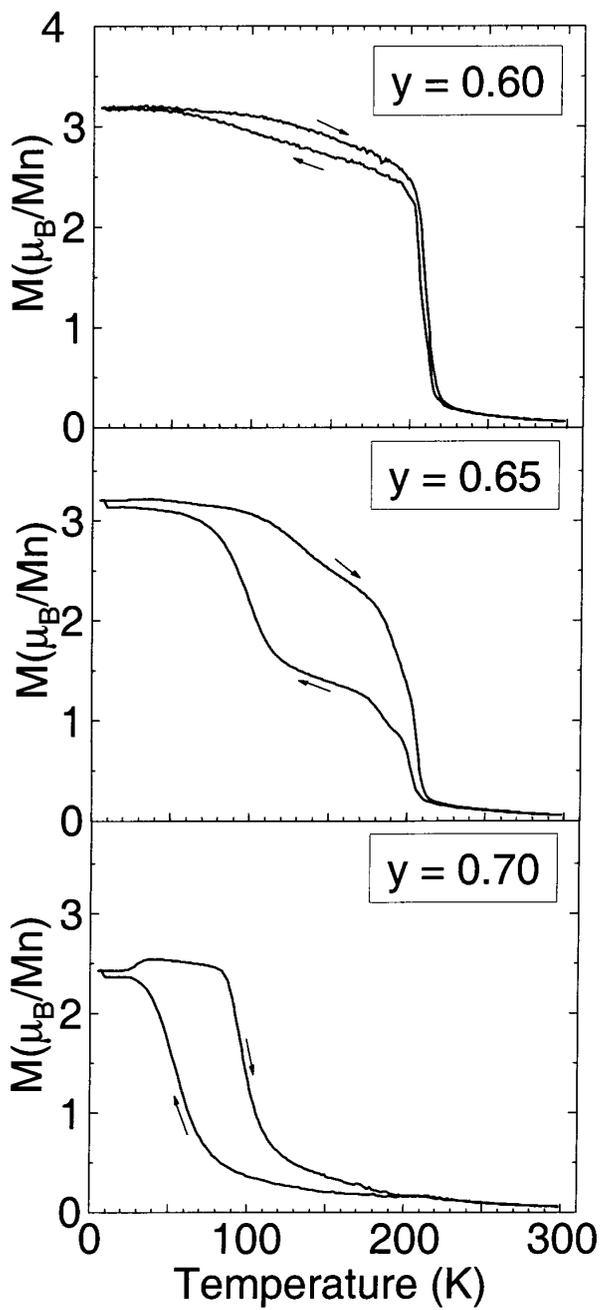

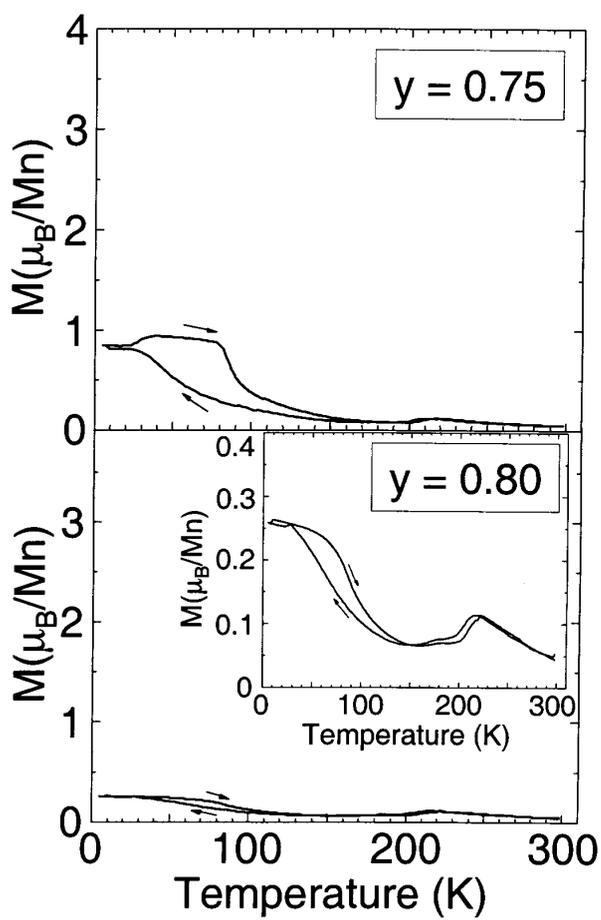

G. Blake *et al.*
Figure 8

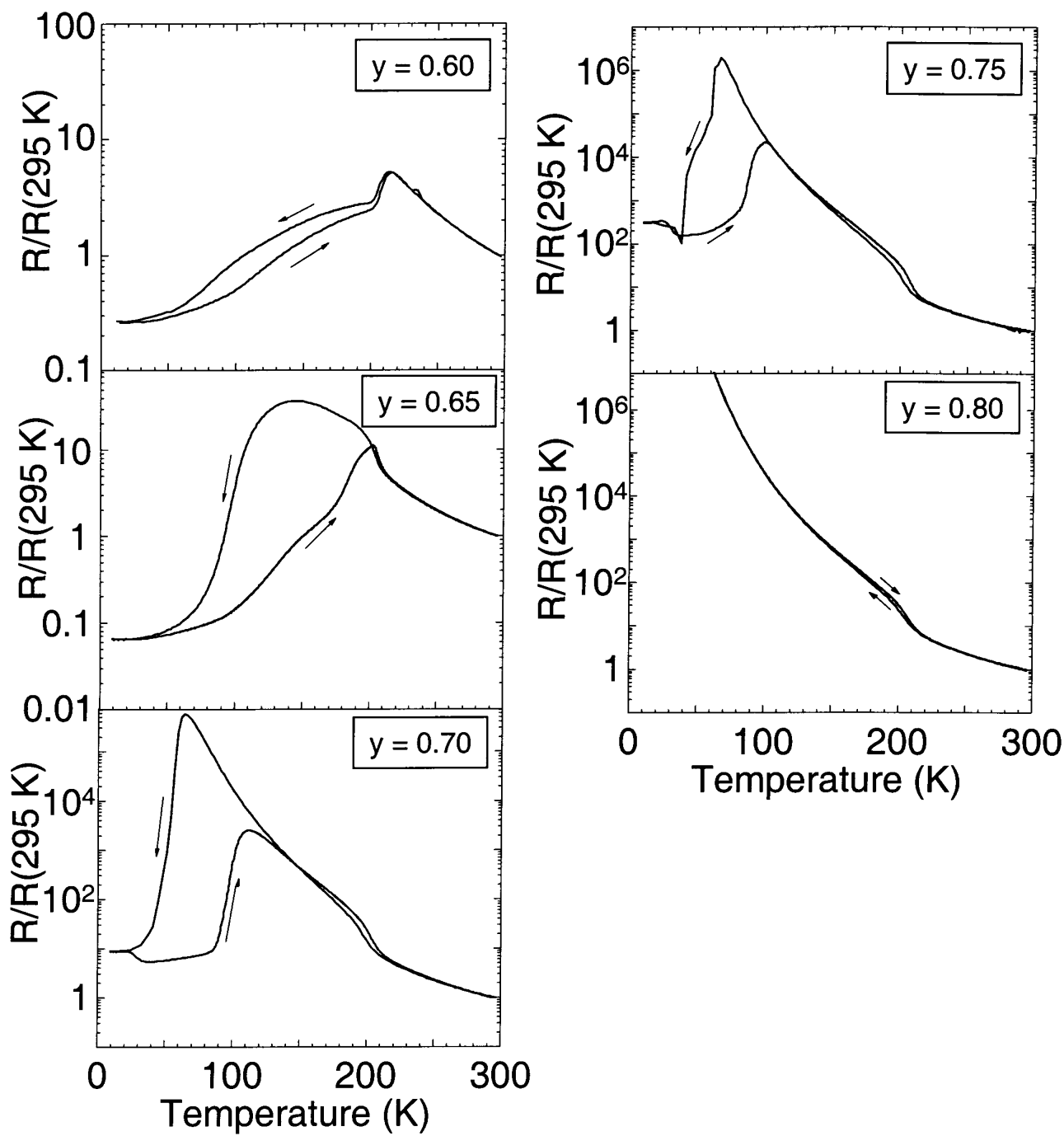

G. Blake *et al.*
Figure 9

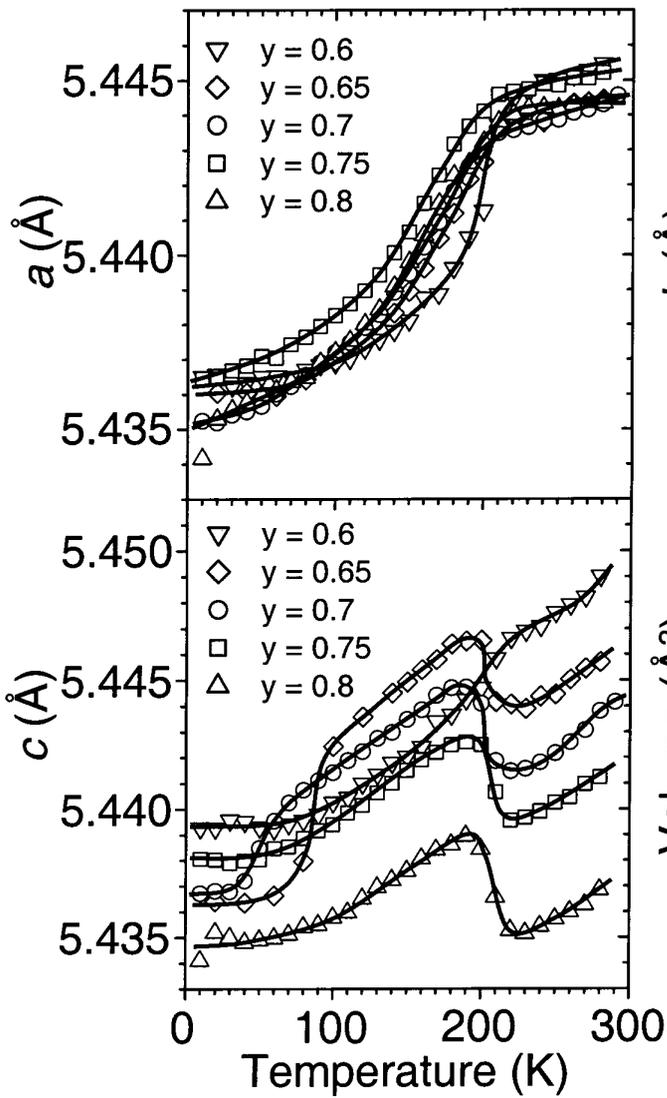
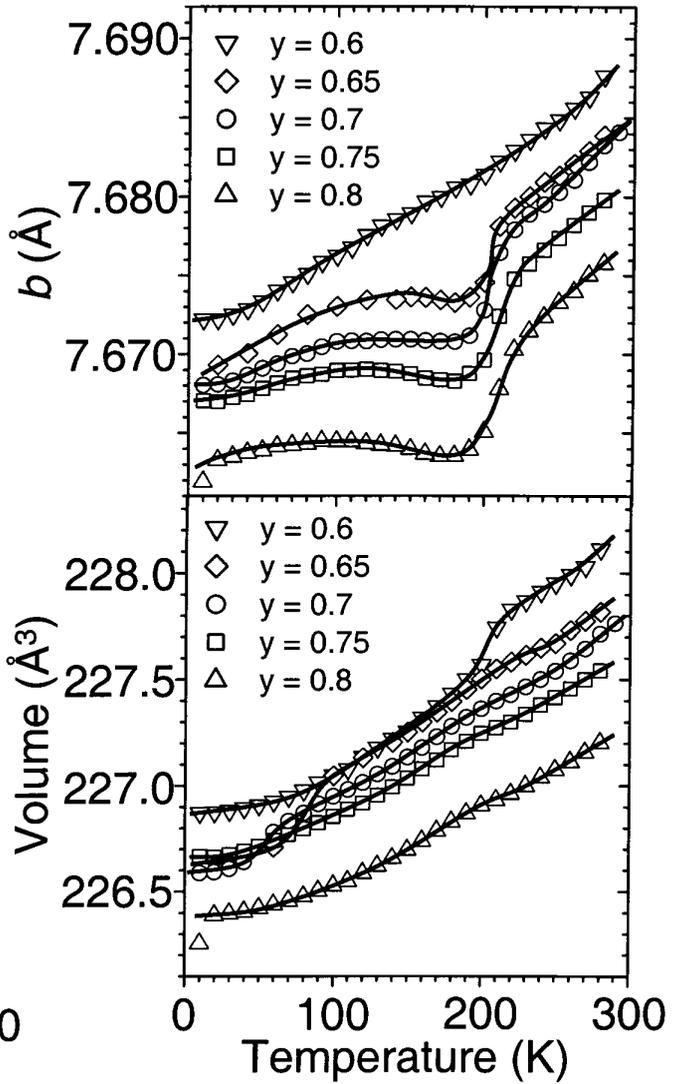

G. Blake *et al.*
Figure 10

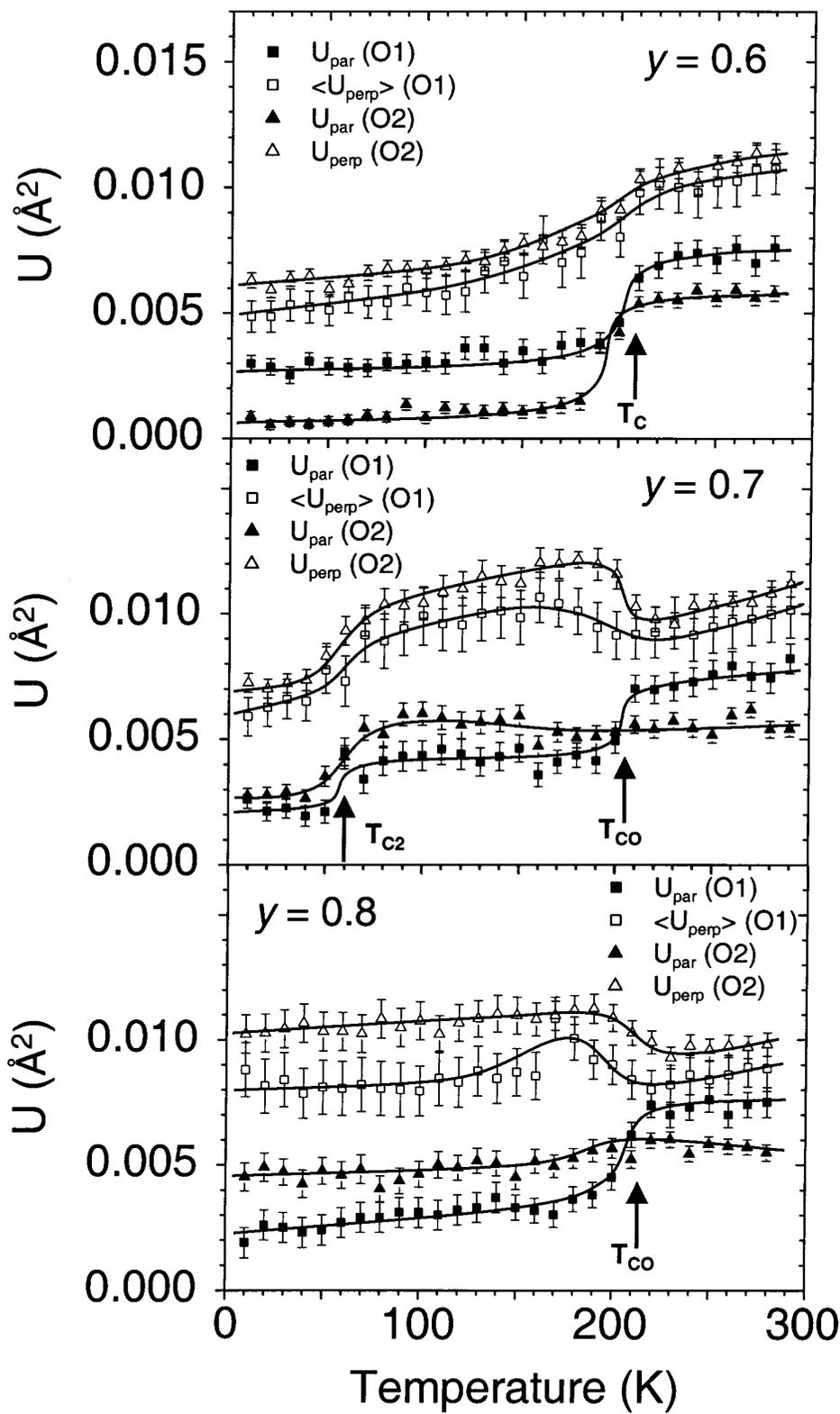

G. Blake *et al.*
Figure 11